\documentclass{iaus}
\usepackage{graphicx}

\title[Molecules in outflows] 
{Molecules in bipolar outflows}

\author[Mario Tafalla \& Rafael Bachiller]   
{Mario Tafalla$^1$
 \and Rafael Bachiller$^1$}

\affiliation{$^1$Observatorio Astron\'omico Nacional (IGN) \\  Alfonso XII 3,
28014 Madrid, Spain \\ email: {\tt m.tafalla@oan.es, r.bachiller@oan.es}}

\pubyear{2011}
\volume{280}  
\pagerange{}
\jname{The Molecular Universe}
\editors{Jos\'e Cernicharo \& Rafael Bachiller, eds.}
\begin{document}

\maketitle

\begin{abstract}
Bipolar outflows constitute some of the best laboratories to study
shock chemistry in the interstellar medium. A number of molecular
species have their abundance enhanced by several orders of magnitude
in the outflow gas,
likely as a combined result of dust mantle disruption and high temperature
gas chemistry, and therefore become sensitive indicators of the
physical changes taking place in the shock. Identifying these species
and understanding their chemical behavior is therefore of high interest
both to chemical studies and to our understanding of the star-formation
process. Here we review some of the recent progress in the study of the 
molecular composition of bipolar outflows, with emphasis in the tracers
most relevant for shock chemistry. As we discuss, there has been rapid progress 
both in characterizing the molecular composition of certain outflows as 
well as in modeling the chemical processes likely involved. However, a number of 
limitations still affect our understanding of outflow chemistry. These
include a very limited statistical approach in the observations and a dependence of
the models on plane-parallel shocks, which cannot reproduce the observed
wing morphology of the lines. We finish our contribution by discussing the
chemistry of the so-called extremely high velocity component, which seems
different from the rest of the outflow and may originate in the wind
from the very vicinity of the protostar.
\end{abstract}

\firstsection 

\section{Introduction}

Bipolar outflows are one of the most studied phenomena
of the star-formation process.
They result from the supersonic acceleration 
of gas in two opposite directions by a newly formed star, and were first 
identified more than three decades ago with radio observations 
(\cite[Snell, Loren, \& Plambeck 1980]{sne80}).
Since their discovery, bipolar outflows have been identified
 around protostars of nearly all masses, from below the brown
dwarf limit to the precursors of the ultra-compact HII regions, and in 
environments as different as isolated globules and cluster-forming regions.
This ubiquity of the outflow phenomenon suggest that bipolar outflow
formation is a necessary element of the physics of star formation,
likely associated to the need for the gas to lose angular momentum
in its process of forming a highly compact object (see 
\cite[Arce et al. 2007]{arc07} for a recent review).


Interest on the chemistry of bipolar outflows goes back in time
as far as the interest on their physics. 
Outflow emission originates from ambient gas that has been accelerated
by a supersonic wind, and therefore has been shock-processed.
As a result, the study of outflow chemistry is unavoidably
intertwined to the study of shock chemistry in the interstellar gas.
Early questions on this chemistry were raised as soon as high velocity 
molecular gas was observed, and were
related to the survival or shock-production of the
CO molecules seen at velocities of tens of km~s$^{-1}$
in the wings of outflow spectra (\cite[Kwan \& Scoville 1976]{kwa76})
and the up to a hundred km~s$^{-1}$ velocities seen in some water masers
(\cite[Morris 1976]{mor76}). 
Prompted by these observations, the first (and still relevant) models
of molecule formation in shocks soon 
appeared (\cite[Hollenbach \& McKee 1979]{hol79}).

\begin{figure}[t]
\begin{center}
 \includegraphics[width=4.5in]{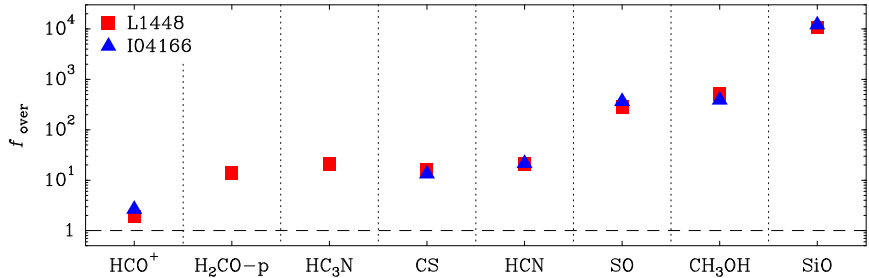} 
\caption{Overabundance factors with respect to dense core values 
for a number of molecules towards the outflows from L1448-mm 
and IRAS 04166+2706. Data from \cite[Tafalla et al. (2010)]{taf10}.
}
   \label{overabundances}
\end{center}
\end{figure}

With time, additional chemical processes have been
identified in the gas accelerated by outflows.
Current interest includes the enhancement in the abundance
of high-density gas tracers like SiO and CH$_3$OH, which is seen
toward a number of outflow sources powered both by high and low-mass
young stellar objects (\cite[Bachiller 1996]{bac96}). 
These abundance enhancements are believed to
result from the erosion of dust grains via sputtering and grain-grain
collisions, perhaps combined with a number of gas-phase chemical 
reactions favored by the temperature increase caused by the 
shock (see \cite[van Dishoeck \& Blake 1998]{van98} for a review).

The amount of the 
abundance enhancement depends on the species under consideration,
and typically ranges from factors of a few for species like HCO$^+$ to several
orders of magnitude for species like CH$_3$OH and SiO (Fig.\ref{overabundances}). 
Not all outflows show the same degree of molecular richness, however,
and this apparent selectivity of outflow chemistry has lead
to the creation of special category of outflows, the so-called ``chemically
active'' outflows (\cite[Bachiller et al. 2001]{bac01}).

\section{Chemically active outflows}

Although easily distinguishable when observed in
species like CH$_3$OH and SiO, the chemically active outflows do not stand 
apart from the rest when observed in CO (the standard outflow tracer),
either by their line strength, velocity extent, or spatial collimation.
It is well established 
that the chemically active phase corresponds to an early period
in the outflow development, as the driving sources of this family of
outflows tend to be Class 0 sources. The exact physical cause of the 
chemical richness is however unclear. One possibility is that the
youngest outflows must encounter denser envelopes along their paths, and 
therefore must naturally give rise to stronger shocks. Another possibility
is that early chemical activity is associated with the higher energy
known to characterize the earliest outflow phases 
(\cite[Bontemps et al. 1996]{bon96}).
In either way, the lack of chemical richness in the more evolved outflows
indicates that chemical activity is highly transient, and that any signature
of its presence must disappear  quickly in the protostellar life
(near the transition between Class 0 and Class I, 
\cite[Tafalla et al. 2000]{taf00}).
Rapid depletion of the enhanced species via freeze out 
onto cold dust grains is the most likely cause of this  effect 
(\cite[Bergin et al. 1998]{ber98}).

\begin{figure}
\begin{center}
 \resizebox{12cm}{!}{\includegraphics{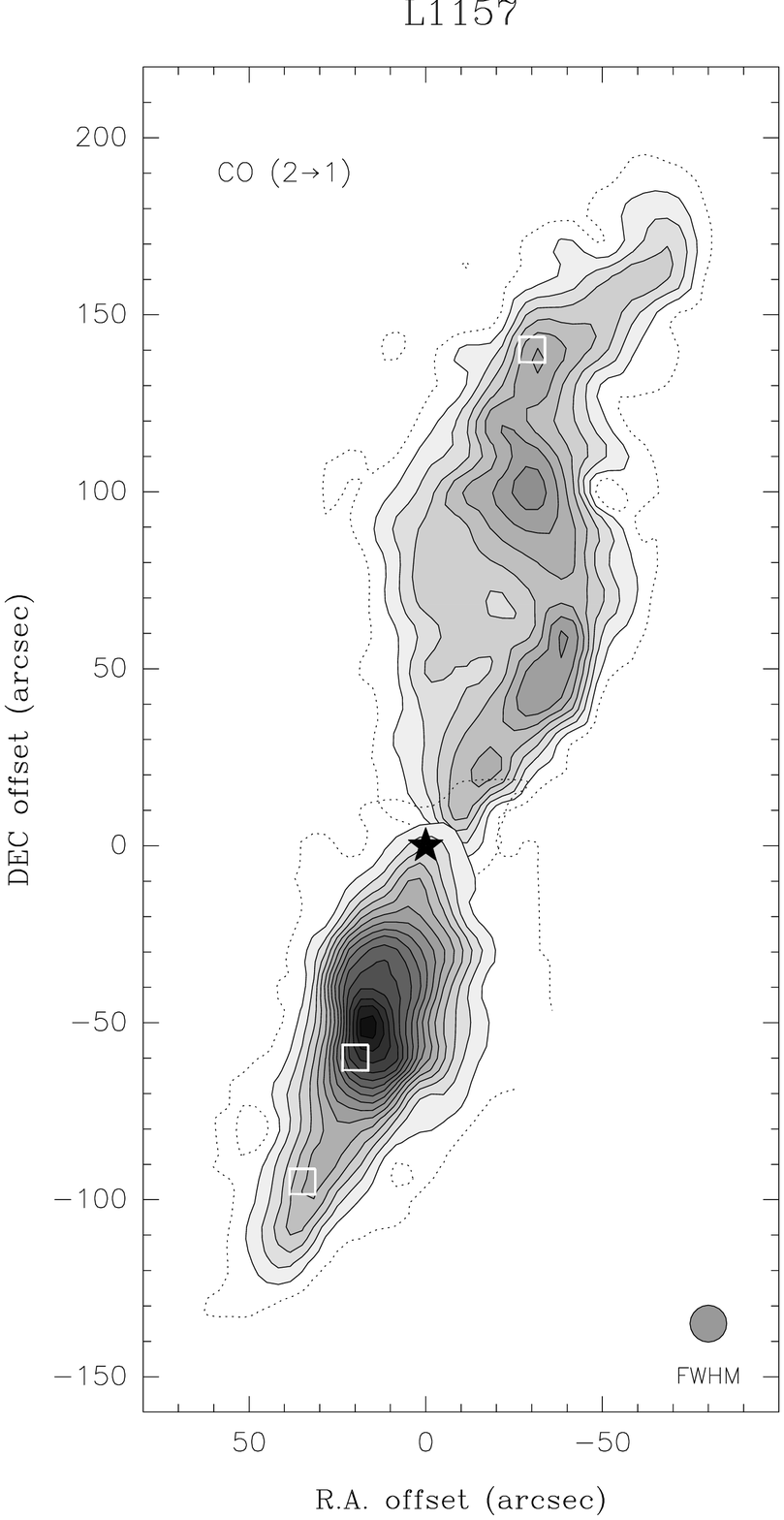}
 \includegraphics{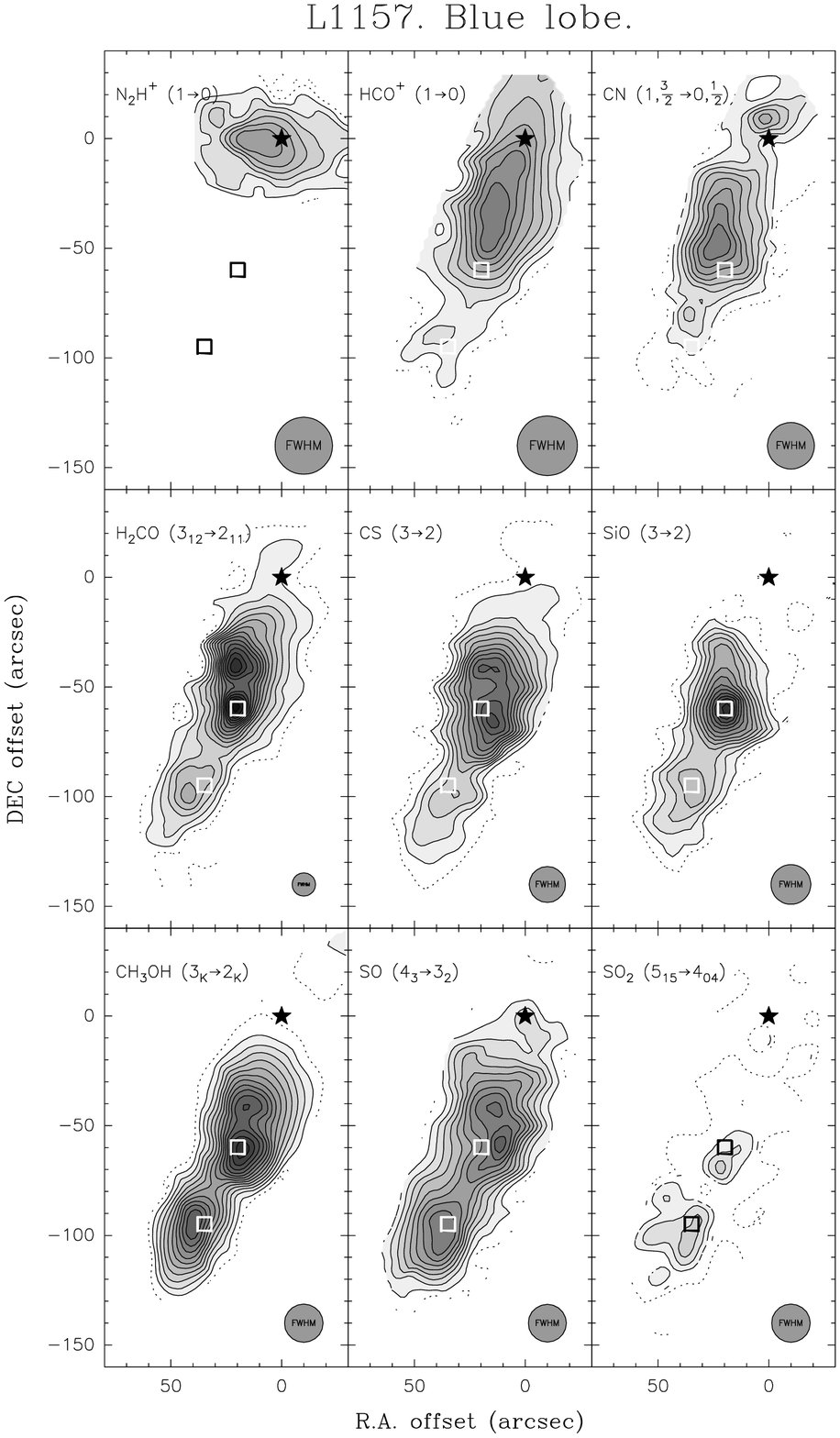}}
\caption{{\bf Left: } map of CO(2-1) emission from the L1157 outflow.
{\bf Right: } maps of integrated intensity for different molecular
species towards the blue (southern) lobe of the L1157 outflow. Note 
the different extension and peak position of each species. 
Figures from \cite[Bachiller et al. (2001)]{bac01}.
}
   \label{l1157_plot}
\end{center}
\end{figure}

Among the group of chemically active outflows, the one in the 
L1157 dark cloud stands
out for the strength of its lines in species like SiO and CH$_3$OH.
As a result, it has been the subject of a very intense observing campaign,
including the study using the Herschel Space Observatory 
presented by Lefloch (2011, this volume).
The L1157 outflow was first identified by \cite[Umemoto et al. (1992)]{ume92} 
due to its strong CO lines. Its chemical activity
was recognized very early on (\cite[Mikami et al. 1992]{mik92}), 
and has been characterized in great detail by 
\cite[Bachiller \& P\'erez-Guti\'errez (1997)]{bac97} and
\cite[Bachiller et al. (2001)]{bac01}.
The chemical activity of L1157 is specially prominent 
towards several regions often denoted as  B1 and B2
(for blue lobe) and R (for red lobe), and are most likely associated 
with regions where the ambient cloud is being strongly shocked by the outflow
wind.  Interferometric 
observations of these regions reveal a complex pattern
of spatial distributions that  indicate a very fragmented 
small-scale structure (\cite[Benedettini et al. 2007]{ben07}). 
As illustration of this chemical richness, we show in
Fig.\ref{l1157_plot} (right panel) a series of maps of the
southern lobe of the L1157 outflow in a number of molecules
that are selectively enhanced in different parts of the outflow 
(especially B1 and B2).  

\section{Recent progress trancing outflow gas with molecules}

It is not possible to review here with detail
the rapid progress made in the study of molecules in outflows 
during the last few years. In this section we present a selection of
studies that have been carried out since the last IAU Astrochemistry
meeting in 2005. We concentrate on low-mass outflow studies and in molecules
that seem specially relevant to our understanding of shock chemistry.

\subsection{H$_2$O}

Probably the most significant progress since the Asilomar meeting 
concerns the
study of the H$_2$O molecule, thanks to the coming on line of a series of
satellite observatories: the Submillimeter Wave Astronomy Observatory
(SWAS), Odin, the Spitzer Space Telescope, 
and the Herschel Space Observatory.

Since the early discovery of its maser emission in outflows,
water has been considered a sensitive outflow tracer. The maser
emission is so bright that it can be easily detected across the galaxy,
and the abundance of water required to produce the emission 
must have been  significantly
enhanced by shocks (\cite[Elitzur \& de Jong 1978]{eli78},
\cite[Kaufman \& Neufeld 1996]{kau96}, \cite[Cernicharo et al.1996]{cer96}). 
The thermal emission of
water, on the other hand, has taken longer to become a recognized
tracer of outflow activity, especially in low-mass outflows. 
Early observations with the Infrared Space Observatory (ISO)
revealed thermal water emission from a number of
bipolar outflows (\cite[Liseau et al. 1996]{lis96},
\cite[Nisini et al. 1999]{nis99},
\cite[Giannini et al. 2001]{gia01}). These ISO spectra, however,
did not resolve spectrally even the broadest outflow profiles,
a fact that was later overcome by the 
SWAS and Odin telescopes, which observed the H$_2$O(1$_{10}$--1$_{01}$) transition
with a resolution better than 1~km~s$^{-1}$.
These instruments, however, had a 
limited angular resolution, approximately 4 arcminutes
for SWAS and half that size for Odin, that only allowed to 
study global averages of the water emission over the whole outflow,
or at least over each of their lobes.
Compilations of this work have been presented by  
\cite[Franklin et al. (2008)]{fra08} (SWAS) and
\cite[Bjerkeli et al. (2009)]{bje09} (Odin).

In the past two years, the Herschel Space Observatory (HSO) has
started to provide water data with  spatial resolution comparable to
that of the large radio telescopes from the ground and, thanks
to its Heterodyne Instrument for the Far Infrared (HIFI), also with
velocity resolutions below 1~km~s$^{-1}$. Two HSO Guaranteed Time Key
Programmes have a strong component of low-mass outflow studies, the
Chemical Herschel Surveys of Star Forming Regions (CHESS) and Water in
Star-forming regions with Herschel (WISH, see 
\cite[van Dishoeck et al. 2011]{van11}), while  
Herschel/HIFI Observations of EXtraOrdinary Sources (HEXOS) 
covers Orion. As a number of contributions in this volume by members
of these teams detail the first results of these programs,
here we will only present a brief description of the new exciting
outflow science that the HSO is providing, and we refer the reader to
the contributions by Bertrand Lefloch (CHESS), Lars Kristensen (WISH),
and Nathan Crockett (HEXOS) for further details.

\begin{figure}[t]
\begin{center}
 \includegraphics[width=4.5in]{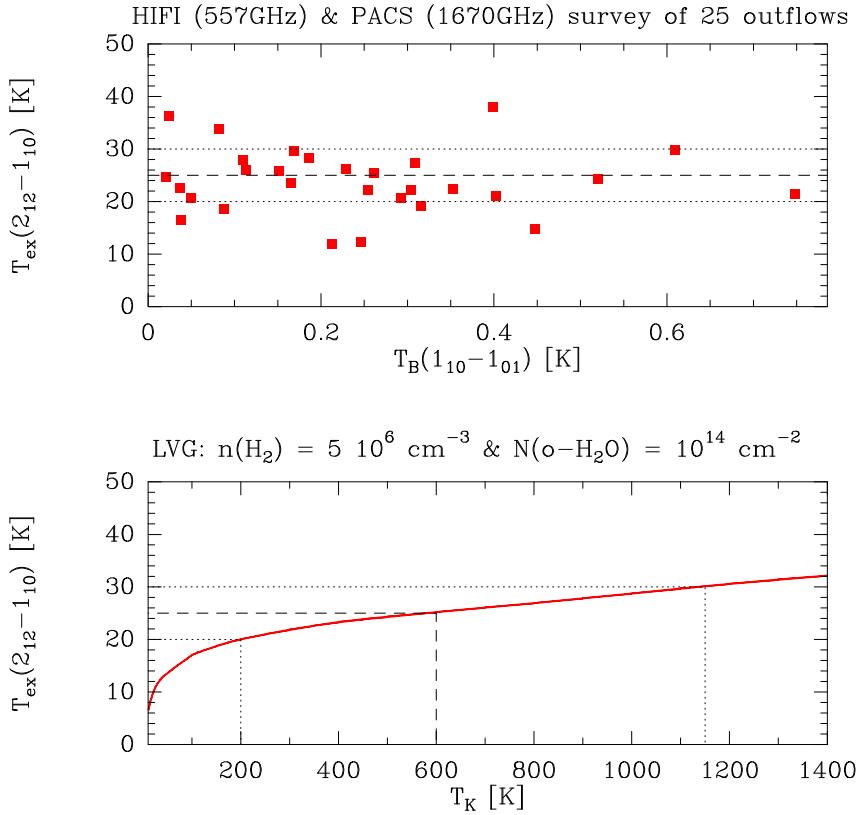} 
\caption{{\bf Top: } excitation temperature between the  first two excited 
levels of o-H$_2$O as a function of peak intensity in the 557~GHz
H$_2$O line, as measured in a sample of 25 outflows within the
WISH program. Note the constant value at a level of  about 25~K 
(optically thin conditions have been assumed). 
{\bf Bottom: } LVG radiative transfer results 
for representative physical conditions of the emitting gas
(n(H$_2$) = $5\; 10^6$~cm$^{-3}$ and N(o-H$_2$O) = $10^{14}$~cm$^{-2}$).
As can be seen, the excitation temperatures derived in the top panel
require gas kinetic temperatures in the range 200-1100~K. 
From Tafalla et al. (2011), in preparation.
}
   \label{wish}
\end{center}
\end{figure}

It is safe to say that we are still exploring the tip of the iceberg
in our research of water emission from outflows. Much work still needs
to be done to understand the excitation conditions of the H$_2$O
emitting gas, and this understanding is critical to extract the
full potential of the water data. Preliminary work on
water excitation in outflows is presented in this meeting by the
poster contributions of Vasta et al. and Santangelo et al., which summarize
the first results of a  multi-transition analysis of water towards the
very young outflows in L1157 and L1448. A more limited treatment in
terms of water lines, but more extensive in terms of outflow sources,
comes from the so-called ``water outflow survey'' carried out also as
part of the WISH project, and illustrated in Fig.~\ref{wish}. This outflow
survey has explored about 25 different objects by observing in each of
them two positions using two ortho-water lines, the 1$_{10}$--1$_{01}$
at 557 GHz  with HIFI and the 2$_{12}$--1$_{01}$ at 1670 GHz with
PACS.  As seen in Fig.~\ref{wish}, the data from 
this survey shows a clear trend for the ratio of
these two lines (when convolved to the same angular resolution) to be
approximately constant independent on the  intensity of the 557 GHz
line. This ratio, assuming optically thin conditions (tested with an
LVG code), is proportional to the excitation temperature between the
first two excited levels of  ortho-water
(T$_{\mathrm{ex}}(2_{12}-1_{10})$), and as Fig.~\ref{wish} (top panel) shows is
approximately constant over the whole sample at a level of
approximately $25 \pm 5$~K. Such a low level of scatter suggests that
the bulk of the water emission we observe in outflows (at least from
the lowest energy levels) arises from a rather narrow range of
physical conditions.

Although  a T$_{\mathrm{ex}}(2_{12}-1_{10})$ value of around 25~K may
seem a relatively low temperature, it is important to remember that the large
Einstein A coefficients of water make this molecule hard to become excited
collisionally. To estimate the kinetic temperature of the gas
responsible for this T$_{\mathrm{ex}}$ of 25~K, we need to estimate
the balance between collisional excitation and radiative de-excitation
(both spontaneous and induced). We have done this using the large
velocity gradient (LVG) approximation together with the recently
calculated collision rates by  \cite[Faure et al. (2007)]{fau07} 
(as provided by the LAMDA data base, \cite[Sch{\"o}ier et al. 2005]{sch05}), 
and assuming values for the gas volume density and H$_2$O
column densities in line with those found by multi-transition analysis
like those in the posters by Vasta et al. and Santangelo et al. As
shown in Fig.~\ref{wish} (bottom panel), T$_{\mathrm{ex}}(2_{12}-1_{10})$ 
values of around 25~K require relatively high gas temperatures, in the
vicinity of 500~K. This result adds to existing evidence that the gas
responsible for the observed water emission from outflows corresponds
to a component warmer than that typically observed with low-J CO
transitions. 

\subsection{H$_2$}

H$_2$ emission from outflows has been observed widely since its discovery
in Orion, and reveals the presence of gas heated to temperatures up to a few thousand
Kelvin  (\cite[Gautier et al. 1976]{gau76}). Most of the H$_2$ work has so far
concentrated on the v=1--0 S(1) line at 2.12 $\mu$m, which is 
easily observable from the ground. A testimony of the widespread nature 
of this emission in outflows is its
use to define the recently created category of 
Molecular Hydrogen emission-line Objects (MHOs), which are analogous to
the optically-based Herbig-Haro objects but include more deeply embedded
line-emitting regions
thanks to its use of NIR wavelengths  (\cite[Davis et al. 2010]{dav10}).
At the time of writing this article, the MHO on-line catalog
({\tt http://www.jach.hawaii.edu/UKIRT/MHCat/}) is growing steadily and
contains more than 1200 entries.

Although some of the pure rotational transitions of H$_2$ are also observable from
the ground (e.g.,  \cite[Burton et al. 1989]{bur89}), sampling the full H$_2$ spectrum,
and therefore characterizing the population of its energy levels, requires the
use of space-based telescopes. ISO observations in the late 1990's carried out the
first complete observations of the H$_2$ spectrum in Orion,
and revealed a distribution of excitation temperatures that 
ranged from about 600~K in the lowest levels to more than 3000~K
in the highest levels (\cite[Rosenthal et al. 2000]{ros00}). 

More recently, the Infrared Spectrograph (IRS) on board of the 
Spitzer Space Telescope has covered the 5-33 $\mu$m 
wavelength range, and therefore has provided access to the
lowest  pure rotational lines of H$_2$.
The high sensitivity of this
instrument has allowed the systematic mapping of 
rotational emission from low mass outflows.
\cite[Neufeld et al. (2006)]{neu06},
for example, observed
the S(0)-S(7) transitions towards the HH 7-11 and HH 54 outflows, and found that
the ortho-to-para ratio is spatially variable, and presents
values significantly below the expected equilibrium value
of 3 for the observed gas temperature (which ranges between 400 and 1200~K).
Such non-equilibrium ortho-to-para ratios can be understood as being a remnant
of the gas pre-shock conditions, where the gas temperature was significantly lower
($\sim 50$~K).

A more systematic study of the H$_2$ emission from low mass outflows using Spitzer
observations comes from the series of papers by
\cite[Neufeld et al. (2009)]{neu09} and \cite[Nisini et al. (2010)]{nis10},
and the poster by Giannini et al. in this meeting, which concentrate on the
outflows from L1157, L1448, BHR71, NGC 2071, and VLA 1623. Despite their
differences in evolutionary stage and central source luminosity, all these
outflows present a number of similar trends. One of them is the need for
a multiplicity of temperatures to explain the H$_2$ emission, and which these
authors fit with a distribution of mass with temperature that follows a
power law $T^{-b}$ with $b$ in the range 2.3-3.3. In addition, the ortho-to-para
ratio in all objects with the exception of L1448 is below the equilibrium
value of 3, and suggest the presence of an activation energy in the para-to-ortho
conversion. Similar results were obtained by 
\cite[Maret et al. (2009)]{mar09} in their
study of the outflows in the NGC 1333 region.

\subsection{Complex Molecules}

There has also been substantial progress in the study of "complex molecules"
in outflows over the last few years.
These organic species are usually defined as containing 6 or
more atoms, and are found in a number of different environments that range from the
Orion hot core to the cold core of TMC-1 (see
\cite[Herbst \& van Dishoeck 2009]{her09} for a recent review).

In outflows, the most widely observed complex molecule is methanol (CH$_3$OH), whose 
abundance is often enhanced by several orders of magnitude, most likely due to its 
release from dust grain mantles (see Fig.~\ref{overabundances}). 
A recent result concerning methanol
comes from the study of its high energy transitions  
by \cite[Codella et al. (2010)]{cod10}, 
who used the Herschel Space Telescope to observe
L1157-B1 as part of the CHESS project. These authors found that
the excitation of the high-J lines (up to J=13) is characterized by a 
rotation temperature of 106~K, which is almost one order of magnitude higher
than the rotation temperature of 
12~K derived from the low-J lines 
(\cite[Bachiller et al. 1995]{bac95}). Given
the excitation conditions of CH$_3$OH, a rotational temperature of more than
100~K implies a gas kinetic temperature of at least 200~K, and such a warm component
represents a link between the low-excitation gas seen in most molecular tracers
and the hotter gas traced by H$_2$ emission. (See also 
\cite[Nisini et al. 2007]{nis07} for a similar
result obtained from SiO observations.)

Molecules larger than CH$_3$OH have also been recently 
identified in L1157-B1. 
\cite[Arce et al. (2008)]{arc08} carried out deep
integrations with the IRAM 30m telescope towards this position 
and detected for the first time in an outflow 
HCOOCH$_3$ (methyl formate), CH$_3$CN (methyl cyanide), 
HCOOH (formic acid), and C$_2$H$_5$OH (ethanol). The detection
of these molecules shows that outflow shocks can lead to a
very rich chemistry, that in low-mass sources was previously restricted
to the hot corino regions in the vicinity of some protostars
like IRAS 16293-2422 (\cite[Cazaux et al. 2003]{caz03}).

Additional detection of complex molecules in the L1157-B1 region comes from a
3mm spectral line survey being carried out 
with the Nobeyama 45m telescope, and whose first results
have been presented by 
\cite[Sugimura et al. (2011)]{sug11}
. This survey has already produced
new detections of complex molecules in this chemically active outflow:
CH$_3$CHO (acetaldehyde) and the mono-deuterated variety of methanol
CH$_2$DOH. From the first analysis of the 
line survey, \cite[Sugimura et al. (2011)]{sug11} 
find that the relative abundance of the complex molecules
with respect to methanol is significantly lower in L1157 than in the prototype hot
corino IRAS 16293-2422. These authors suggest that such a difference reveals a 
difference in the chemical processing of the dust in the cloud regions sampled by
the outflow shock compared to the vicinity of the protostar sampled in the hot corino
phase.  Detection of more complex molecules is possible given the on-going nature of the 
survey.

Another recent detection towards the L1157-B1 position is that of HNCO
(isocyanic acid) by 
\cite[Rodr{\'{\i}}guez-Fern{\'a}ndez et al. (2010)]{rod10}.
Although not a
complex molecule, HNCO is a well-known organic tracer of warm
environments, like hot cores, the Galactic center, and even
extragalactic sources, and the good correlation of its distribution
with that of methanol in IC 342 had led to the suggestion that this
species is shock-sensitive (\cite[Meier \& Turner 2005]{mei05}). 
The detection of
HNCO in L1157-B1 by \cite[Rodr{\'{\i}}guez-Fern{\'a}ndez et al. (2010)]{rod10}
not only
reinforces this suggestion, but illustrates the important role of
relatively simple regions like L1157-B1, which is far enough from the
YSO to be considered a ``pure shock'' region (without stellar heating),
and therefore provides a clean template to recognize shock chemistry
in more complex environments.

\section{Do we understand outflow chemistry?}

A meeting like this one is an appropriate occasion  to look critically at the
state of our understanding of outflow chemistry in a more global way than
usually done in  research papers. A look of the literature reveals a
generalized consensus in the belief that shock chemistry can explain most (or all)
abundance anomalies observed in the outflow gas (apart from the
EHV component discussed in the following section). This consensus, however, seems
based more on the lack of observational counter-examples and alternative
models than in the existence of a quantitative proof that a single shock chemistry 
model can explain simultaneously the
set of abundances observed in an object like L1157-B1. 

This incomplete state of affairs mostly results from a 
too-focused approach in outflow-chemistry work. From the
observational point of view, and this paper testifies it, most of this work 
has been dedicated to a single object, the L1157 outflow, and more exactly, to 
the B1 position of this outflow. Although L1157 is sometimes called ``prototypical''
in terms of outflow chemistry,
the lack of objects with comparable chemical richness and complexity 
suggest that L1157 is probably extreme,
if not for its abundances, at least for the column densities responsible for its
very bright lines.
 
\begin{figure}[t]
\begin{center}
\includegraphics[width=2.0in, angle=-90]{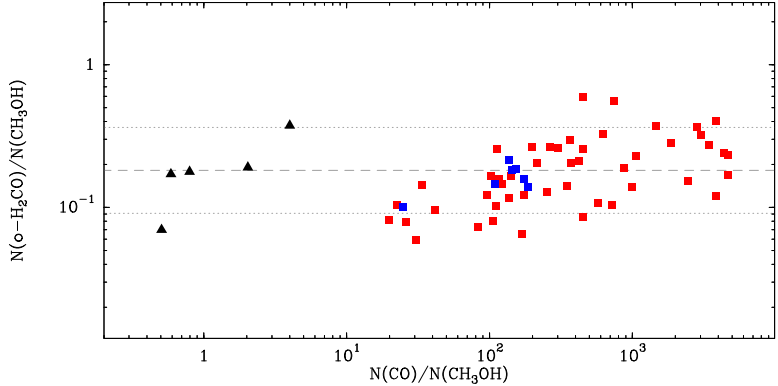}
\caption{Abundance ratio between H$_2$CO and CH$_3$OH in the outflow gas phase
(squares) and in the ices of dust grains (triangles). The red squares 
represent values for a
sample of low-mass outflows from Santiago-Garc\'{\i}a et al. 
(in preparation), and the blue squares are values towards high-mass
star forming regions from
\cite[Bisschop et al. 2007]{bis07}. The black triangles are ice
values from \cite[Gibb et al.(2004)]{gib04}.
The good match between gas-phase and solid-phase values reinforces the idea
that the abundance enhancement of H$_2$CO and CH$_3$OH in outflows results from
the release of these molecules from the mantles.
}
   \label{h2co_ch3oh}
\end{center}
\end{figure}

Clearly, more general studies of outflow chemistry are needed, not only 
to understand how typical L1157 is, but to reconstruct the full cycle of outflow chemical
activity represented by other very young outflows.
An important limitation for any statistical approach is the lack of 
good samples. This is partly the result of the small number of outflows that
so far have been explored in molecular lines other than CO, but also  because of the 
need for full maps to identify chemical hot spots in the flows. Despite the limitations, 
some work has been done in this direction. 
Fig.~\ref{h2co_ch3oh} shows an example of how an statistical 
approach can further illuminate the origin of the
observed abundances. This plot represents the CH$_3$OH/H$_2$CO
ratio for the sample of low-mass
outflows studied by Santiago-Garc\'{\i}a et al. (in preparation), and
shows that this gas-phase ratio remains approximately constant over a 
relatively large range of individual abundances. The (gas-phase) ratio, 
in addition, matches
closely the solid-phase ratio derived from ice mantles
in a different sample of objects, and this good match 
supports strongly the interpretation that the
CH$_3$OH and H$_2$CO abundance enhancements in outflows result from the
release of these species from the mantles of grains due to sputtering in a shock.

While observations tend to concentrate on one object at a time, chemical models tend 
to solve one 
species at a time, and  this makes it difficult to asses whether a single set of physical
conditions can explain the variety of observations, even for a single object like L1157. 
A more important limitation
of the chemical models is their reliance on plane-parallel shock models. 
In a plane parallel shock, there is a single velocity for the shock, and all
the material suffers the same type of acceleration. As a result, the predicted 
emerging spectrum is characterized by a single spike at the velocity of
the shock and a rapidly declining tail of pre-shocked material towards low velocities 
(e.g., Figs. 8 in both \cite[Gusdorf et al. 2008]{gus08} and 
\cite[Flower \& Pineau Des For{\^e}ts 2010]{flo10}).
This shape of the spectrum is the opposite to what it is observed in a typical
outflow wing, which has most of the material moving at low velocities 
and a minority of the gas moving at top speeds (see next
section for a discussion of the different nature of the 
spiky extremely high velocity regime). As the optical depth of the 
emission is proportional to the amount of material per unit velocity,  
the compression of shocked gas into a small
range of velocities predicted by the plane parallel model 
leads to a substantial overestimate of the optical depth.
By comparing in the figures mentioned above
the width of the spikes with the full velocity extent of the spectrum, we estimate 
that the  optical depth overestimate resulting from the artificial 
limited range of velocities in the emission must range from factors of a few to
more than order of magnitude. Thus, comparing observations
with the predictions from chemical models must be done with care when the 
model predicts a non-negligible of optical depth in the emission.

The systematic reliance on plane-parallel models in the analysis of 
outflows is more a reflection of our 
poor understanding of the internal kinematics in the outflow gas
than a conviction in the model providing the correct geometry.
In fact,
the observation of a wing feature in the emission spectrum of almost 
every outflow indicates that a multiple-velocity description is needed, 
and that most of the gas in the outflow
moves at intrinsically low velocities.
How to incorporate this description to chemical models is
not yet clear, although a first step may be to parametrize the mix
of velocities
with a power-law description like that used in the multi-temperature
analysis of the
H$_2$ data by  \cite[Neufeld et al. (2006)]{neu06} and described above. 
Using such a more realistic description of the outflow velocity 
field will clearly bring a better agreement between the predicted
and observed spectra. More interestingly, a multiple-velocity
description of the shocked gas will allow exploring the changes in
the gas composition as a function of the velocity that a number of
observations are starting to uncover 
(\cite[Codella et al. 2010]{cod10}, \cite[Tafalla et al. 2010]{taf10}).

\section{The EHV gas, a different chemical component?}

\begin{figure}[t]
\begin{center}
 \includegraphics[width=3.5in]{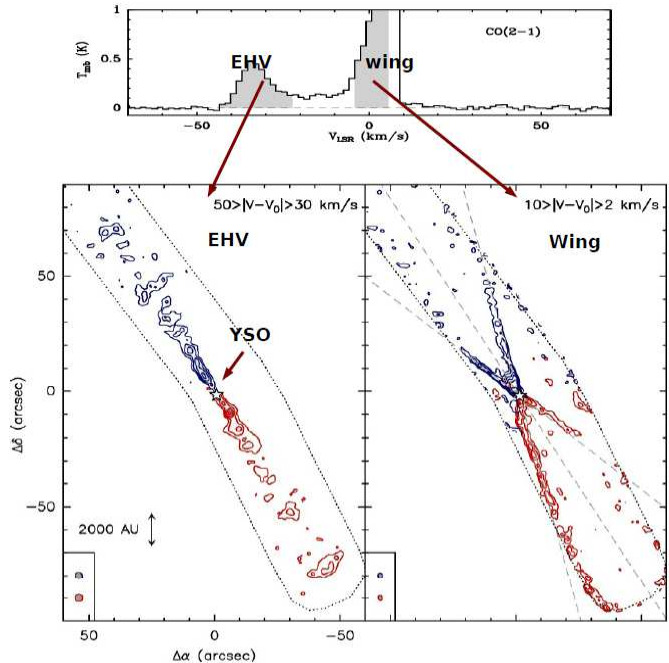} 
\caption{Average blue lobe spectrum and velocity-integrated maps
of the CO(2-1) emission from the IRAS 04166+2706 outflow illustrating the
different spatial distribution and spectral signature of the wing and the 
EHV components. Data from Santiago-Garc\'{\i}a et al. (2009).
}
   \label{i04166_ehv}
\end{center}
\end{figure}

So far we have referred to the ``outflow emission'' as the one appearing in the spectra
forming a wing where the intensity decreases steadily with velocity. 
Although such type of wing emission is the most characteristic signature of outflow gas, a
few very young Class 0 objects show in their spectra an additional outflow
component that looks like a close-to-gaussian secondary peak. This component, often referred to
as the extremely high velocity (EHV) gas given its high speed (tens of km~s$^{-1}$), 
was  first discovered in the outflow powered by L1448-mm 
(\cite[Bachiller et al. 1990]{bac90}), and by now it has
been observed in a (reduced) number of outflows. 

One of the outflows where an EHV component has been recently identified is
the one powered by IRAS 04166+2706 (IRAS 04166 hereafter, see
\cite[Tafalla et al. 2004]{taf04}).
In Fig.~\ref{i04166_ehv} we present the CO(2-1) emission from this outflow 
as observed by 
\cite[Santiago-Garc{\'{\i}}a et  al. (2009)]{san09}
using the Plateau de Bure Interferometer. The
figure shows that the EHV emission appears distinct from the 
wing emission both in spectral signature and spatial distribution. As can be seen, the
(lower-velocity) wing component forms in the maps 
an X-shaped structure centered on the IRAS source and suggestive of
tracing the walls of a pair of evacuated cavities (a fact confirmed by the
association of the northern blue cavity with an IR cometary nebula).
The EHV emission, on the other hand, appears in the maps as a jet-like
feature that runs along the middle of the cavities and consists of a collection
of discrete peaks. As shown by 
\cite[Santiago-Garc{\'{\i}}a et  al. (2009)]{san09}, the
peaks of EHV emission are located so symmetrically from the IRAS source
that each has a counterpart on the other side
less than $2''$ away from its expected position.
Such a level of symmetry suggests that the EHV peaks
arise from
events that took place near the central source
and that have since propagated outwards
with the flow.

\begin{figure}[t]
\begin{center}
\resizebox{12cm}{!}{\includegraphics[width=4in,clip]{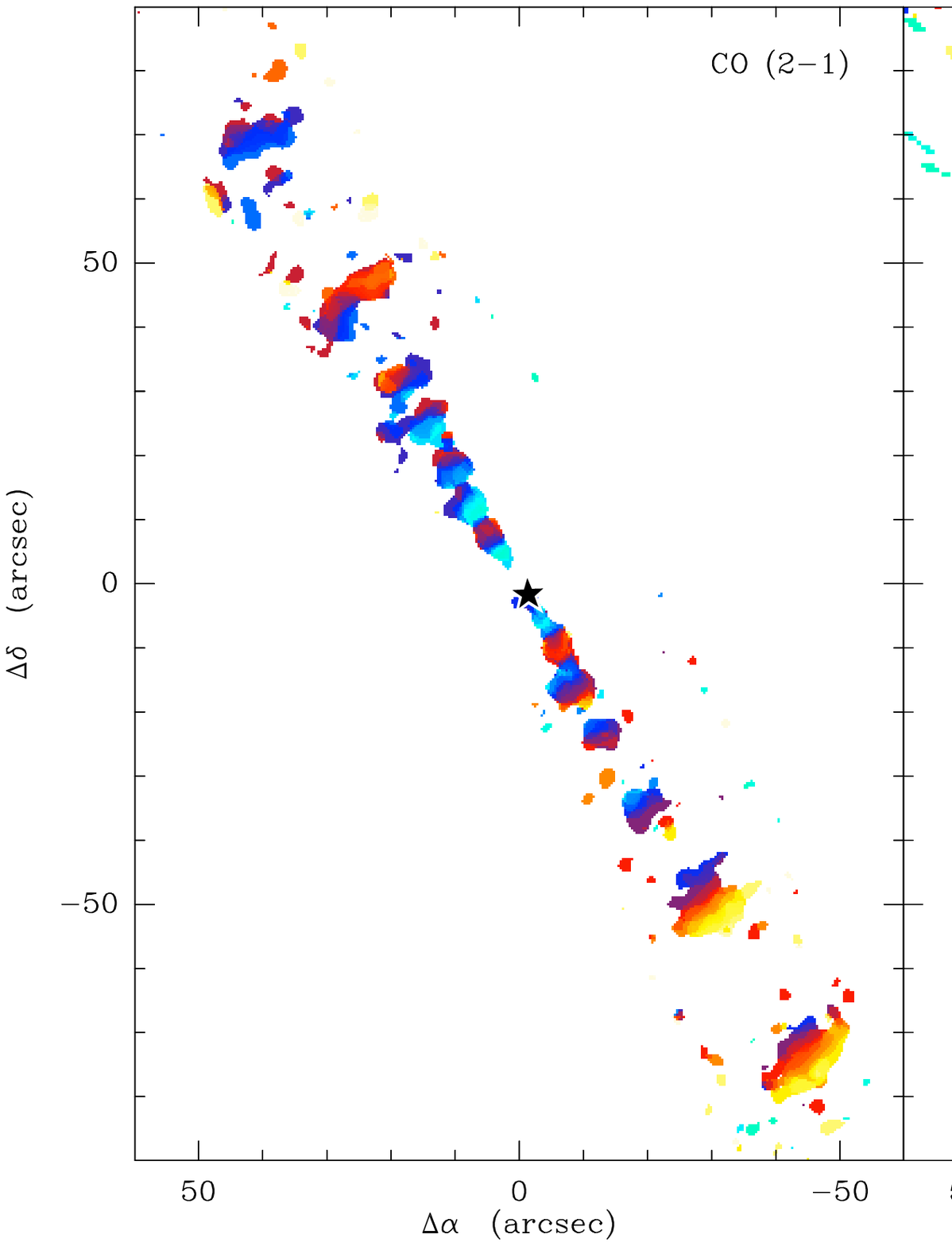}
 \includegraphics{i04166_co_pv_2.ps}}
\caption{{\bf Left:\ } First momentum plot of the CO(2-1) EHV 
emission towards the outflow
from IRAS 04166+2706 illustrating the presence of velocity oscillations.
{\bf Right:\ } Position-velocity diagrams of CO(2-1) along the outflow axis.
Note the sawtooth pattern in the EHV regime. From 
\cite[Santiago-Garc\'{\i}a et al. (2009)]{san09}. 
}
   \label{i04166_vel}
\end{center}
\end{figure}

A clue to the origin of the EHV peaks
comes from a pattern of oscillations in their
velocity field.
As illustrated in Fig.~\ref{i04166_vel},
the oscillations form a sawtooth pattern in the
position-velocity diagram indicative of a combination of a close-to-constant
mean velocity together with strong velocity gradients inside each EHV peak.
The sense of these gradients is the same in all EHV peaks, and indicates
that the upstream gas moves faster than
the downstream gas. This sawtooth velocity pattern in the PV diagram 
is in striking agreement
with the predictions for a set of internal shocks in a jet
modeled by \cite[Stone \& Norman (1993)]{sto93} to simulate the
pattern of knots seen in optical jets (see their Figure 16).
Thus, the combination of a symmetric and 
fragmented structure together with a sawtooth velocity
pattern suggests that the EHV emission in IRAS 04166 
arises not from ambient accelerated gas (like the rest of the
outflow), but from the internal shocks in a pulsating jet.

If the EHV gas represents jet material, it must be coming from the protostar
or its near-most vicinity, and we can therefore expect its chemical composition to
differ substantially from that of the wing gas, which consists
of ambient gas that has been shock-accelerated, and therefore has 
had a very different thermal history. To test this possible difference
between these two components of the outflow,
\cite[Tafalla et al. (2010)]{taf10} have carried out the first molecular survey
of EHV gas by making deep integrations towards two outflow positions
known to have bright EHV and wing components, one in the L1448 outflow
and the other in the IRAS 04166 outflow. Previous to this survey, only
CO and SiO had been detected in the EHV gas of any outflow,
despite the large number of species detected in the wing component
of outflows like that of L1157 (see above). Thanks to the new deep
integrations, carried out with the IRAM 30m telescope, and some of them lasting 
several hours, the number of species observed in the EHV gas has more than
doubled, with clear detections of SO, CH$_3$OH, and H$_2$CO. Possible
detections of HCO$^+$ and CS have also resulted from these deep integrations,
although their status remains unclear because of possible contamination
with emission from the wing outflow component, which extends at a low
level up to velocities comparable to those of the EHV gas.
(See also \cite[Kristensen et al. 2011]{kri11}  for
the recent detection of H$_2$O in the EHV gas of L1448.)

\begin{figure}[t]
\begin{center}
 \includegraphics[width=4.5in]{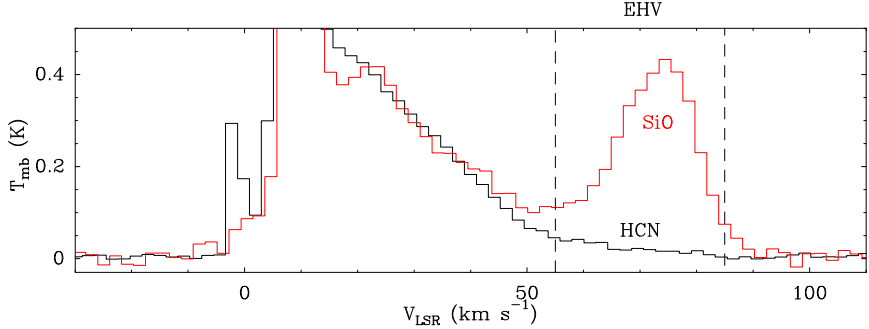} 
\caption{Comparison between SiO(2-1) and HCN(1-0) spectra towards 
the red lobe of the L1448-mm outflow. The SiO line has
been scaled to match the HCN intensity in the wing, in order
to better illustrate the sudden drop of HCN/SiO
abundance in the EHV component. From \cite[Tafalla et al. (2010)]{taf10}.}
   \label{l1448_hcn_sio}
\end{center}
\end{figure}

Probably the most interesting result from the EHV molecular survey of
\cite[Tafalla et al. (2010)]{taf10} is the evidence for a change in the
carbon-to-oxygen ratio between the wing and the EHV components.
This is illustrated in Fig.~\ref{l1448_hcn_sio} with a superposition of the
SiO(2--0) and HCN(1-0) spectra towards the target position in the
L1448 outflow  (the IRAS 04166 target position presents
a similar behavior but weaker lines). 
As can be seen, when scaled appropriately, the  
SiO(2-1) and HCN(1-0) spectra track each other in the (red) wing 
regime over several tens of km~s$^{-1}$, suggesting that the HCN/SiO
abundance ratio remains constant over this range of velocities.
In contrast with this smooth behavior in the wing regime,
the intensity ratio between HCN and SiO drops by more
than one order of magnitude in the EHV regime. This intensity drop 
is unlikely to result from a difference in excitation between the 
two velocity
regimes, as multi-transition analysis of SiO, SO, and CH$_3$OH
show only small changes in the excitation temperature of these
molecules between the wing and EHV gas. The most likely cause of
the sudden drop in the HCN intensity towards the EHV regime is a
similar, order-of-magnitude drop in the abundance of this species
towards this outflow component. A similar drop in the abundance of
CS is also seen in the data, suggesting 
that the HCN drop is not a peculiarity of this molecule but a common
feature of C-bearing species. Indeed, all molecules with clear detection in
the EHV component have so far been O-bearing (CO, SiO, SO, CH$_3$OH, 
H$_2$CO, and H$_2$O).

The finding of chemical differences between the EHV and wing outflow components
in both L1448 and IRAS 04166 
opens a new tool to explore the still mysterious mechanism of outflow
acceleration. If the EHV gas truly arises from a jet, it must have
originated in the innermost vicinity of the central object, and
therefore must carry information about the physical conditions in the
outflow acceleration region, which is of a few AU or even
a few stellar radii, depending on the model, and therefore
inaccessible to current instrumentation. To fully exploit the
potential of this information, however, a new generation of
chemical models is needed. The chemistry of protostellar
winds was originally studied by 
\cite[Glassgold et al. (1991)]{gla91},
but little progress has been done on this issue in the twenty
years past since this pioneering work. It is reassuring to see
that there is a renewed interest in the topic, as illustrated 
by the poster contribution in this meeting of Yvart et al.
Clearly more modeling and observational progress is needed
in this new and exciting line of study of molecules in outflows.
Let's hope we can all see its results in the next IAU Astrochemistry
Symposium a few years from now.


\begin{discussion}

\discuss{Sternberg}{Could you comment on whether it might be possible to study
angular momentum loss using these molecular diagnostics?}

\discuss{Tafalla}{Rotation in molecular jets has been searched for, and there are
a few possible detections in the literature. Our observations on the jet from
IRAS 04166, however, suggest that the kinematics of the gas in the 
jet is dominated by 
lateral ejection in the internal shocks, and I suspect that this faster
motion is going to overwhelm any rotation signature that may be present.}

\discuss{Ellinger}{You have shown similar abundances of HCN in three different objects.
Do you observe the same thing for HNC?}

\discuss{Tafalla}{Bachiller \& P\'erez Guti\'errez (1997) estimated an HNC/HCN 
ratio of 0.1 towards L1157-B1. In the fast wing of L1448, the non detection
of HNC in Tafalla et al. (2010) sets an upper limit that is about one order of 
magnitude lower.}

\discuss{Menten}{What do we know about deuterium enhancement in the outflow gas,
for example for methanol?}

\discuss{Tafalla}{Sugimura et al. (2011) have recently presented
the first detection of mono deuterated methanol towards L1157-B1, and derive an
CH$_2$DOH/CH$_3$OH abundance ratio in the range 0.013-0.029, which is significantly
lower than that derived for IRAS 16293 but similar to the one measured in Orion KL.}

\discuss{Visser}{You showed that the HCN/SiO abundance ratio is much lower in the EHV 
component than it is in the wing, and you interpreted this as a lower elemental 
C/O ratio in the EHV gas. Could the low HCN/SiO ratio also indicate a low elemental
N/O ratio?}

\discuss{Tafalla}{It could be. Still, we find a similar drop in the CS/SiO ratio
between the wing and the EHV gas, although we have lower signal-to-noise in the CS
data. This seems to point more to a deficit of carbon than of nitrogen, although
we are dealing here with a very limited set of observations.}

\end{discussion}

\end{document}